\documentclass[acmsmall, nonacm]{acmart}

\AtBeginDocument{%
  \providecommand\BibTeX{{%
    \normalfont B\kern-0.5em{\scshape i\kern-0.25em b}\kern-0.8em\TeX}}}

\setcopyright{rightsretained}
\copyrightyear{2021}
\acmYear{2021}
\acmDOI{10.1145/1122445.1122456}

\acmJournal{JACM}
\acmVolume{37}
\acmNumber{4}
\acmArticle{111}
\acmMonth{8}

\begin{document}

\title{Toward a Theory of Programming Language and Reasoning Assistant Design: Minimizing Cognitive Load}

\author{Michael Coblenz}
\email{mcoblenz@umd.edu}
\orcid{0000-0002-9369-4069}
\affiliation{%
  \institution{University of Maryland}
  \streetaddress{8125 Paint Branch Drive}
  \city{College Park}
  \state{Maryland}
  \country{USA}
  \postcode{20742}
}


\newcommand{\anonymous}[1]{(anonymous for review)}

\begin{abstract}
	Current approaches to making programming languages and reasoning assistants more effective for people focus on leveraging feedback from users and on evaluating the success of particular techniques. These approaches, although helpful, may not result in systems that are as usable as possible, and may not lead to general design principles. This paper advocates for leveraging theories from cognitive science, focusing on cognitive load theory, to design more effective programming languages and reasoning assistants. Development of these theories may enable designers to create more effective programming languages and reasoning assistants at lower cost.
\end{abstract}



\maketitle

\section{Introduction}


How can we design programming languages and reasoning assistants that make users as effective as possible? User-centered design techniques, which have recently been applied to programming languages~\cite{Coblenz2020:Usercentered}, enable designers to iteratively improve their designs using feedback from users. Alternatively, when a designer has a design candidate that they believe may be effective, the designer can evaluate the tradeoffs in a user study. However, there are two problems with these approaches. First, perhaps the designs are beneficial relative to other design candidates, but alternative designs might be superior. That is, perhaps the process finds only a \textit{local maximum} in the design space. Second, the approach does not provide \textit{predictive power} regarding candidate designs, and in a search for effective designs, it is not cost-effective to evaluate large numbers of design candidates. The results of evaluating individual systems may not generalize to new systems that combine techniques from earlier evaluated designs because of potential interactions among design components. Of course, in particular situations, the designer might \textit{hypothesize} that prior results \textit{do} apply on the basis of the designer's experience or on the basis of various kinds of theory.

An alternative approach to design languages that are more effective for users would be to use a theory of usability of languages. If we had theories that provided predictive power regarding programming language and reasoning assistant (PL/RA) designs, designers might be able to create designs that are more effective overall, avoiding the \textit{local maximum} challenges that arise with user-centered design iteration and the \textit{high costs} of repeated evaluations with users. Even if a theory does not provide quantitative predictions, it may help a designer hypothesize which designs are likely to be more effective than others, guiding the design process at lower cost than an empirically-driven approach.

To illustrate the benefits of a user-centered theory-driven approach to PL/RA design, in this paper, I consider the relevance of one particular theory (\textit{cognitive load theory}, CLT) to programming language designs. Many of these implications pertain to type systems or the design of reasoning assistants directly, though for the sake of completeness, I consider other implications as well. In the future, we might turn our efforts to developing \textit{new} theories, potentially in collaboration with cognitive scientists. By showing how one  theory from cognitive science may have implications on PL/RA design, I aim to promote the use and further development of human-related theory in PL/RA design. These theories complement traditional mathematical theories of programming languages, which should be used simultaneously. By coupling an analysis of the cognitive load characteristics of a feature and participants' task performance in user studies, it may be possible to develop a predictive model of how designs affect language learnability and programming task performance.

\section{Introduction to Cognitive Load Theory}

Cognitive scientists have long modeled human memory as including a \textit{short-term memory} component, which temporarily stores facts that are relevant to one's current task~\cite{Glanzer1966:Two}. Unfortunately, the capacity of short-term memory is limited, with researchers finding limits between 4 and 7 items in adults~\cite{Cowan2010:Magical}. In addition, items in short-term memory can be discarded as a result of a change in attention or due to the passage of time. Cognitive load theory (CLT) was developed by Sweller~\cite{Sweller1994:Some} in a context of studying learning. CLT models learning challenge in terms of the number of items that one must remember at once in order to accomplish the learning objective. If the learning process requires remembering too many items, then performance suffers in terms of time required, errors made, or both. CLT applies to problem-solving tasks, particularly ones conducted in the context of learning~\cite{Sweller1988:Cognitive}.

How can we write software effectively, considering the complexity of software (even a single function or method likely has more than 4 lines of code)? There are three techniques that allow us to work around the limits of short-term memory. First, some relevant facts can be located in \textit{long-term memory}. Although adding items to long-term memory requires rehearsal, long-term memory has an extremely large capacity. Second, by combining disparate facts into one entity (``chunking''), the facts can share one memory ``slot.'' For example, perhaps beacons~\cite{Wiedenbeck1986:Beacons} reflect a way to chunk lines or tokens so that they require less short-term memory. Finally, one can externalize information by recording information in another form, such as in a text file or on paper. Of course, paging this externalized information back in has a time cost, and may be neglected entirely. Failure to restore externalized information could be a source of bugs, as important requirements or facts are omitted from consideration.

When cognitive load is high, CLT predicts lower learning efficiency~\cite{Sweller1988:Cognitive}. High cognitive load is also associated with high error rates~\cite{Ayres2001:Systematic}. Even in small lab studies with 30-40 lines of code, cognitive load has been found to be high when certain antipatterns are followed in identifier selection~\cite{Fakhoury2018:Effect}. Existing methods (also described by Fakhoury et al.~\cite{Fakhoury2018:Effect}) can be used to measure cognitive load, allowing researchers to investigate the effect of PL/RA designs on cognitive load in relevant tasks.

CLT also predicts an effect called \textit{expertise reversal}~\cite{Kalyuga2003:Expertise} in which providing cognitive support directed at novices can actually reduce experts' performance on the same tasks. The idea is that experts may experience unnecessary cognitive load when processing beginner-level advice that they do not need. This might point toward a need for customized error messages or other tools. 

\section{Impact of Designs on Cognitive Load}

Programming languages, environments, and proof assistants each embody design choices that impact their users' cognitive load. Key techniques for reducing cognitive load include reducing the amount of information needed for a task (as in features that promote modularity); co-locating relevant information (typically provided by IDE features such as \emph{jump to definition}; and visualizations. 

\subsection{Programming Languages}

What programming language design decisions might influence short-term memory requirements? Any programming subtask that does not directly relate to one of the high-level requirements may consume short-term memory slots. Below is a partial list of programming tasks that may unnecessarily occupy short-term memory.

\begin{description}
\item[Determining reference validity] When programming in a language in which references can be invalid, reference validity may need to be checked. If neglected, this can result in undefined behavior (in unsafe languages) or runtime errors.

\item[Leveraging imprecise static type information] Types constrain the operations that are permitted on an expression. When using an unfamiliar type, a programmer might read declarations or documentation to determine what operations are available, so these operations are in short-term memory, not long-term memory. As a result, doing a sequence of operations with the type might require re-reading the documentation. IDEs can help via autocomplete tools, but these tools are less effective when an expression has an imprecise type, since less information is available regarding what operations are available.

In addition, autocomplete and autosuggest features can be made much more precise in the presence of precise static type information; these features prevent users from needing to remember the full names of uncommonly used methods and functions. 

\item[Handling errors] Handling errors requires that the programmer consider additional questions that do not pertain to the programmer's primary goal, such as: what might have caused the error? Is the error recoverable? What should be done about it? C programmers are expected to check return values for exit codes, but many of the errors are not recoverable (for example, there is typically no solution if \texttt{malloc} returns \texttt{NULL}). Java programmers need to handle exceptions, but frequently do not bother to write appropriate handlers~\cite{Kery2016:Examining}. Future research should investigate whether writing error handling code is problematic from a cognitive load perspective and consider whether separating handling errors out as a separate development task might reduce cognitive load.

\item[Implicit operations] Should a language allow implicit casts? If so, the semantics of operations may be unclear; if not, the need to insert explicit casts may increase cognitive load (since the programmer must think about the choices of which types to cast to and from). Some languages, such as Scala and Agda, allow implicit arguments. This may reduce cognitive load if the arguments are not relevant to the task, but it may increase it if the programmer needs to figure out how the implicit arguments are inferred.

\item[Reasoning about code re-use] Changing one function or module may break dependent modules. Therefore, when making a change, the programmer should be aware of any requirements imposed by clients of the changed code. This requires that the programmer add the dependencies to short-term memory; keeping all of these in mind simultaneously could easily exceed the size limit. One might argue for clear module boundaries and complete requirement specifications, but specifications can be redundant with parts of the implementation (must one re-state preconditions that are expressed in assertions?), and specifications can themselves be inconsistent with the code.

PL designers typically provide features for \textit{modularity} to make it easier to re-use code in different contexts and to protect clients of code from changes in other parts of the codebase. Although this separation of concerns can reduce cognitive load, it can come at a cost: more abstract code can result in more abstract or less-localized error messages, which can be harder to understand. In addition, leveraging modularity features results in more architectural complexity, which can itself increase cognitive load for clients. That is, there can be a tradeoff between load for clients and load for API authors. One approach could be to use program optimization or partial evaluation~\cite{Blazy1998:Partial, Harman1997:Amorphous} to reduce information needs for authors of clients. \footnote{Thanks to reviewer 1 for this observation.}. 

Inheritance is one language feature that results in hidden dependencies; the author of a superclass can be unaware of subclasses (the fragile base class problem~\cite{Mikhajlov1998:Study}). Type systems are a key tool designers can use to promote and inhibit code re-use, in part through subtyping. Future work might investigate how subtyping impacts cognitive load.

\item[Managing effects] Designers of pure languages, such as Haskell, often tout the cognitive benefits of avoiding side effects. From a cognitive load perspective, side effects may demand that the programmer keep the possible side effects in short-term memory when calling a function rather than being able to reason only based on the inputs and outputs. If the effects pertain to a module that is not in short-term memory, the programmer may need to load that module into short-term memory, potentially flushing other needed information.

However, pure languages may also impose costs on short-term memory. Programmers may need to track additional state explicitly (e.g., with variables that represent the current environment), requiring additional slots of short-term memory. If the state could have been externalized in a way that does not require programmer attention, a stateful solution may well be better from a cognitive load perspective. By providing a channel of communication among parts of a program that is independent of control flow, state can reduce the need for state transmission to occupy short-term memory slots.

Type systems that capture effects may make the problem even worse by forcing programmers to attend to effects that are irrelevant to their concerns. For example, they may not care whether the function logs to the console, but one cannot do so without specifying an appropriate effect, which may increase cognitive load. On the other hand, effect systems could reduce cognitive load by preventing the programmer from having to consider all possible effects. Future research should evaluate the cognitive load impact of particular effect system designs.

\item[Fulfilling proof obligations] When using some verification tools, users may need to use ghost variables and other techniques to fulfill proof obligations. These variables increase cognitive load relative to approaches that do not require them (including ones that do not use verification).

\end{description}

\subsection{Programming Environments}

\textbf{Requirements and types.} Software components typically need to meet collections of requirements. Functions or methods may have requirements regarding security, performance, invariant maintenance, error checking, error handling, and of course functional correctness. Even this list of six categories may be larger than the short-term memory size! It is not surprising that many functions do not, in fact, meet all of their requirements without extensive testing. Perhaps the process of writing tests helps programmers divide their work in a way that lets them focus on just one or two requirements at a time --- a number that \textit{does} fit in short-term memory. Then, they can fix their implementations by using an external requirement store: the set of tests.

\textit{Requirements traceability} refers to the ability to relate requirements (as typically written in a natural language requirements document) to code or other artifacts that implement the requirements~\cite{IEEETerminology2010}. This is useful in software engineering so that software engineers can answer questions about the purposes of particular sections of code. What if a programming environment provided explicit requirements traceability functionality? The IDE could track pending requirements for each module, and a programmer could annotate components when fulfilling those requirements. Even without any kind of verification process to mechanically ensure the requirements have been met, the mechanism could still serve to help people keep track of which requirements they needed to meet and which they had attended to so far.

Some IDEs provide limited support for tracking unmet requirements by generating warnings that correspond to comments that start with \texttt{TODO}. But this requires that programmers recall unmet requirements while writing code. Instead, suppose programmers first listed requirements, and then worked toward fulfilling them (an approach often recommended by instructors of beginning programming courses!). The process is similar to that proposed by test-driven design or test-first design, but lighter weight because it does not require writing actual test cases first. Instead, prose requirements would suffice, and the IDE could help the programmer track which ones were still to be completed. New maintainers could re-check the revised code against the original list of requirements, and the IDE could help by providing a checklist to examine.

How would the IDE connect code to requirements? Verification approaches, such as dependent types, are one approach, but these carry a significant burden for users. Another approach, when formal verification is not cost-effective, is for the code to carry metadata that maps between natural language requirements and lines of code.

\textbf{Dependency management.} Many languages require that code both import required modules in code (e.g. Java \texttt{import}) as well as specify version of libraries that the program depends on (which the build tool should fetch before building, e.g Maven dependencies). When adding a use of a new library, the programmer must set aside the current task, manage the dependencies, and then return to the current task --- at which time the programmer must remember what was in their short-term memory. Could programming environments help queue this kind of work to avoid overloading the programmer's short-term memory? To what extent must these tools be workflow-specific --- or can one class of management tool apply in many different kinds of situations?

\textbf{Information access.} When working on programming tasks, programmers frequently need a variety of different kinds of information: documentation, code examples, notes, and of course the code being edited. Jump-to-definition features help programmers easily access relevant information for some tasks. Some systems, such as Code Bubbles~\cite{Bragdon2010:Code}, help users arrange some of these items on the screen at once. This mitigates short-term memory limitations, which prevent programmers from keeping all the relevant information in their minds. However, IDEs typically do not provide assistance keeping track of relevant facts \textit{automatically} other than with a back/forward button. Can IDEs automatically detect which relevant pieces of information are unlikely to fit in short-term memory and provide appropriate assistance?

\textbf{Visualizations.} Although architectural diagramming tools exist, automatic visualization tools frequently generate diagrams that are too densely-populated for practical purposes. Programmers may need diagrams that include information relevant to their current tasks, but many visualization tools incorporate only limited task information when they generate the visualizations. Visual representations matter in problem solving, however. For example, Zhang~\cite{Zhang1999:Nature} showed how the representation of Tic-Tac-Toe can make it seem like a completely different game. 

\subsection{Proof Assistants}
Interactive proof assistants, such as Coq~\cite{Coq2021}, typically track the pending proof obligations that the user is required to fulfill. This obviates the programmer from needing to track these in short-term memory. They also track the set of theorems that have already been proven. The latter presents a challenge, however, since in working toward a particular proof obligation, the user must either repeatedly scan the list of available theorems, or commit them to their  (small) short-term memory. Could proof assistants \textit{chunk} the available theorems in order to help the user fit more of them in short-term memory? For example, could they \textit{combine} them where possible to create a smaller number of more-powerful theorems? 

Tactics-based theorem provers require users to learn a set of available tactics that can be applied to transform proof terms. Experts have large sets of tactics available in their long-term memories; novices have learned much smaller sets. But in many cases, the best tactic to apply may not be available in long-term memory and also not have been used recently. Perhaps proof assistants could \textit{suggest} tactics to be applied. Unlike automatic proof search, the idea would be to suggest specific tactics that might be useful, rather than applying a sequence of them automatically. If the suggestions are relevant enough of the time, this approach could help users learn tactics that they could use in the future. Unlike traditional textual autocomplete in IDEs, this form of autosuggest would not be based on lexical prefixes, but instead on proof obligations and available proof terms.

\section{Related Work}

Crichton et al.~\cite{Crichton2021:Role} studied the role of short-term memory in code tracing tasks, finding a limit of about 7 variable/value pairs and proposing changes that could be made to programming tools to help with code tracing tasks. Reducing cognitive load has been proposed in other HCI contexts, such as multimedia interfaces~\cite{Oviatt2006:Human}. Shaffer et al. proposed techniques that might reduce cognitive load in introductory programming courses, such as partitioning information into small segments~\cite{Shaffer2003:Applying}.

Helgesson~\cite{Helgesson2021:Exploring} studied grounded theory analysis of interviews and literature regarding cognitive load in software engineering, finding several categories of contributions to cognitive load:  task, environment, information, tool, communication, interruption, structure and temporal. This paper focuses on the \textit{tool} perspective.

Programming with holes~\cite{Omar2019:Live} and active code completion~\cite{Omar2012:Active} are both approaches that might help programmers maintain focus on their tasks. Programming with holes allows programmers to defer work for later, leaving type information to direct their follow-up work. Active code completion allows domain-specific user interfaces to guide the user through use of a specialized API, preventing the user from needing to consume short-term memory slots with API details. Likewise, Agda~\cite{Bove2009:Brief} has an emacs mode that allows the user to insert typed holes for completion later.

Code Bubbles~\cite{Bragdon2010:Code} and JASPER~\cite{Coblenz2006:JASPER} help programmers collect and reference relevant pieces of information while working in an IDE. Both tools allow programmers to arrange relevant information, such as code, on the screen while working on tasks. These techniques may help reduce programmers' reliance on short-term memory.

\section{Future Work}

The list of design decisions that may impact cognitive load is surely incomplete; a more exhaustive list (and a method for enumerating it) would be preferable. Future work should investigate whether the design decisions described above indeed relate to cognitive load, whether the cognitive load leads to a usability cost, and if so, how the cognitive load might be decreased. Furthermore, cognitive load theory is only one of many existing theories from cognitive science; others, such as the concept of \textit{attention}, should be investigated as well. Finally, the theoretical frameworks from cognitive science are unlikely to suffice to justify every PL/RA design decision, so the community should work toward richer theories that provide more explanatory power regarding programmer behavior.

\section{Conclusion}

Existing theories from cognitive science, such as cognitive load theory, may offer guidance on a variety of programming language design questions. Many of those questions have been the subject of debate or empirical evaluation, but having a theory (particularly one with sufficient predictive power) could lead to a more effective design strategy for programming languages and reasoning assistants. This paper proposes exploration of the relationship between the identified language design decisions, cognitive load, and performance, with the hope of assessing whether cognitive load theory can be leveraged in programming language design.

\begin{acks}
Thanks very much to Clayton Lewis, Daniel Helgesson, and the anonymous HATRA reviewers for their extensive feedback and suggestions on drafts of this paper.

\end{acks}

\bibliographystyle{ACM-Reference-Format}
\bibliography{HATRA-2021.bib}
\end{document}